\documentclass{icrc2009}

\usepackage{graphicx}   
\usepackage[caption=false]{caption}    
\usepackage[font=footnotesize]{subfig} 
\usepackage{fixltx2e}
\usepackage{url}

\newcommand{\shorttitle}[1]%
{\markboth{Proceedings of the 31\MakeLowercase{$^{st}$} ICRC, {\L}\'{o}d\'{z}         2009}{#1} }
\newcommand{\etal}{\MakeLowercase{\textit{et al. }}} 

\usepackage[thinspace]{SIunits}
\usepackage{color}
\usepackage{xspace}
\def\Offline{\mbox{$\overline{\textrm{Off}}$\hspace{.05em}\protect\raisebox{.4ex}
{$\protect\underline{\textrm{line}}$}}\xspace}

\begin{document}
\title{A MC simulation of showers induced by microscopic black holes}
\author{\IEEEauthorblockN{D. G\'ora\IEEEauthorrefmark{1}\IEEEauthorrefmark{2}, M.     Haag\IEEEauthorrefmark{1}
and M. Roth\IEEEauthorrefmark{1}}
\IEEEauthorblockA{\IEEEauthorrefmark{1}Karlsruhe Institute of Technology (KIT),   D-76021 Karlsruhe, Germany}
\IEEEauthorblockA{\IEEEauthorrefmark{2}Institute of Nuclear Physics PAN, ul.   Radzikowskiego 152, 31-342 Krak\'ow, Poland}}

\shorttitle{D. G\'ora \etal A MC simulation of showers induced by microscopic   black-holes}
\maketitle

\begin{abstract}
  Large surface detectors might be sensitive not only with respect to extensive air   showers induced by ultra high energy neutrinos but also to showers induced by   hypothetical objects like microscopic black-holes. Microscopic black-holes might   be produced in high energy particle collisions with the center of mass energies   above the fundamental scale of gravity. These black-holes would decay rapidly by   Hawking radiation into characteristic high multiplicity states of Standard Model   particles and induce extensive air showers potentially detectable by a large   surface neutrino detector.  In this paper we study the possibility to detect   microscopic black-holes exemplifying it in case of the surface detector of the   Pierre Auger Observatory. The expected event rate is calculated for up-going and   down-going showers induced by microscopic black-holes. Our calculations show a   significant deviation of the expected rate compared to the rate expected by   Standard Model predictions. The rates of up-going neutrinos are almost completely   suppressed, whereas the rate of down-going neutrinos increase by a factor of about   50 with respect to standard model calculations. The non observation of up-going   neutrinos by the Pierre Auger Observatory in conjunction with a high rate of   down-going neutrino-induced showers, would be a strong indication of physics   beyond the Standard Model.
\end{abstract}
\section{Introduction}
Searching for ultra high-energy (UHE) neutrinos ($10^{18}\usk\electronvolt$ or above) emitted from astrophysical objects is one of the most challenging tasks in Astroparticle Physics. Neutrinos offer a unique opportunity to open a new observation window, since they are only weakly interacting and neutral. After having traveled cosmological distances without being perturbed and/or deflected in the interstellar medium, neutrinos behave as messengers of the most mysterious regions of astrophysical sources.  Several theoretical models predict a significant flux of high-energy neutrinos as a result of the decay of charged pions, produced in interactions of UHE cosmic rays within the sources themselves or while propagating through background radiation fields.


The expected neutrino rates possibly detected by large surface detectors do not only depend on the predicted flux of neutrinos but also on the neutrino-nucleon cross section.  It has been noted~\cite{Fargion:1997ft} that Earth-skimming $\nu_{\tau}$'s will generate upward going air showers when they interact in the earth crust. By contrast neutrinos of all flavours will generate deeply penetrating quasi-horizontal (down-going)  air showers which are distinctive in having an electromagnetic component unlike hadron induced showers~\cite{Zas:2005zz}. The rate for down-going $\nu$ induced showers is proportional to neutrino-nucleon cross section, while the rate of Earth-skimming $\nu_{\tau}$ is not.  Thus, the detection rate for down-going $\nu$s, and the rate for up-going $\nu$ showers, react differently to variations of the neutrino cross section and tau energy loss. By comparing these rates one can therefore constrain significant deviations of neutrino interactions from SM predictions, see for example~\cite{Anchordoqui:2006ta} for more details.

The neutrino-nucleon cross section is related to parton densities in the yet unmeasured low Bjorken-$x$ region $x \sim 10^{-5}$. Some models even propose substantial modifications of neutrino interactions at high energies, including theories of \tera\electronvolt-gravity and production of microscopic black holes (BH), a domain that is to be tackled in the near future by the LHC.  Measuring the flux of ultra high energy neutrinos would not only allow to put limits on cosmic ray production and propagation models. It would also probe fundamental interactions at energies that lie well above the \tera\electronvolt{}~scale, and open a new window on possible physics beyond the Standard Model (SM)~\cite{Anchordoqui:2006auger}.

In this paper the scenario of microscopic BH production in neutrino nucleon collisions at the \tera\electronvolt{}~scale is exemplified for the case of the surface detector (SD) of the Pierre Auger Observatory~\cite{Abraham:2004dt}, which consists 1600 water Cherenkov detectors with 1.5\usk\kilo\metre{} spacing.  In addition to hadrons and photons the Auger Observatory is also sensitive to UHE neutrinos with energies above $\sim10^{17}\usk\electronvolt$~\cite{Abraham:2008zz,tiffenberg}.

The outline of the paper is as follows. In section II a short description of microscopic BH physics is given. In section III a full MC simulation chain starting with the injection of a neutrino into the Earth's atmosphere, its propagation, interactions and eventually the air shower production up to the actual response of the Auger SD array is described.  Finally, in section IV results of the calculations event rates are presented.

\section{Black hole production}

In conventional 4-dimensional theories the Planck scale
$\sim 10^{19}\usk\giga\electronvolt$ (at which quantum effects of gravity become strong)
is fundamental and the study of black holes lies beyond the realm
of experimental particle physics. In theories where the existence of larger warped extra dimensions is suggested, the 4-dimensional Planck scale is derived from the D-dimensional fundamental Planck scale, which can be of $\mathcal{O}(\tera\electronvolt)$, bringing the possible observation of BH production and evaporation into reach of available experiments~\cite{Randall:1999ee,Giddings:2000ay}.




Microscopic BHs might be produced in high energy particle collisions with the center of mass (CM) energies above the fundamental scale of gravity.  When the impact parameter of two incident particles drops below the Schwarzschild radius $r_\mathrm{s}$ of a BH with mass equal to their CM energy, BH formation should occur with a mass $M_{\mathrm{BH}} = \sqrt{\hat{s}}$.  In case of a neutrino-nucleon collision the squared CM energy $\hat{s}$ is given by $\hat{s} = x \cdot 2 m_{\mathrm{N}} E_{\nu}$, where $m_\mathrm{N}$ is the  nucleon mass and $E_{\nu}$ the energy of the interacting neutrino.  Within a semi-classical approach this suggests a geometric cross section of $\hat{\sigma} \approx \pi r_\mathrm{s}^2 $ with
\begin{equation}
  r_\mathrm{s} \left( M_\mathrm{BH} \right) =   \frac{1}{M_D}\left[\frac{M_\mathrm{BH}}{M_\mathrm{D}}\right]^{\frac{1}{1+n}}\left[\frac{2^n \pi^{\frac{n-3}{2}} \Gamma \left(\frac{n+3}{2}\right)}{n+2}\right]^{\frac{1}{1+n}}
\end{equation}
so that $ \hat{\sigma} \propto \hat{s}^{\frac{1}{n+1}} $, where $n$ is the
number of extra dimensions.

The quantity $M_\mathrm{D}$ denotes the lightest possible BH with a trans-Planckian Schwarzschild-radius.

The total cross section for the formation of a BH in neutrino-nucleon collisions can then be written as
\begin{equation}
  \sigma \left( \nu N \rightarrow BH \right)  = \sum\limits_{i}       \int\limits_{(M_\mathrm{BH}^{\min})^2/s}^{1} dx \, \hat{\sigma_i}\left(\sqrt{xs}\right) f_i       \left( x,Q \right),
\end{equation}
where $f_i(x,Q)$ denotes the parton distributions functions (PDF) and $M_\mathrm{BH}^{\min}$ is the minimal BH mass for which a semi-classical treatment of BH formation is expected to be valid and not understood effects of quantum gravity can be neglected.
Due to the rapidly rising nucleon PDF at low $x$, the effective BH mass $M_\mathrm{BH}$ does not exceed several tens of $\tera\electronvolt/{\rm c}^2$, even for neutrinos with an energy up to $10^{21}\usk\electronvolt$.

Once produced, microscopic BHs are expected to decay within time scales of $\sim 10^{-25}\usk\second$ through three major phases:
\begin{itemize}
\item The balding phase, in which the 'hair' (asymmetry and moments due to the   violent production process) is shed.
\item The Hawking evaporation phase~\cite{Hawking:1974sw}, which consists of a short   spin-down phase (the Kerr (rotating) BH loses its angular momentum) and then a   longer Schwarzschild phase, which accounts for the greatest proportion of mass   loss~\cite{Page:1976df}.
\item A Planck phase at the end, when the BH mass or the Hawking temperature reach   the Planck scale.
\end{itemize}
Given the validity of the semiclassical description, a BH will mainly evaporate due to Hawking radiation and behave like a thermodynamical system with the temperature $T_\mathrm{H} = \frac{n+1}{4 \pi r_\mathrm{s}}$.  During the decay the Hawking temperature will rise as the BH mass drops. The lifetime can be described as $\tau \sim \frac{1}{M_\mathrm{D}} \left(\frac{M_\mathrm{BH}}{M_\mathrm{D}}\right)^{\frac{3+n}{1+n}}$. Approximately the decay may be treated as instantaneous on detector time scales, since for $M_\mathrm{D} \sim1\usk\tera\electronvolt$ and $M_\mathrm{BH} \sim 10\usk\tera\electronvolt$, the BH lifetime $\tau$ is smaller than $10^{-25}$ s. During the decay process particles of all SM channels will be produced in a 'flavor-democratic' fashion with energies of order $T_\mathrm{H}$ or above, typically at relatively high multiplicities $\langle n \rangle \approx \frac{1}{2}\frac{M_\mathrm{BH}}{T_\mathrm{H}}$.


\section{Method}
In order to calculate the expected event rate from microscopic BH-induced showers at SD array of the Auger Observatory, a full MC simulation chain was set.  Simulation consists of three phases: propagation and interaction of neutrinos inside the Earth and atmosphere to produce primaries able to initiate potentially detectable showers in the atmosphere; simulation of lateral profiles of shower developments in the atmosphere and, finally, simulation of detector response.
\begin{figure*}[h]
\centering
\includegraphics[angle=0, width=0.48\textwidth, height=5.99cm]{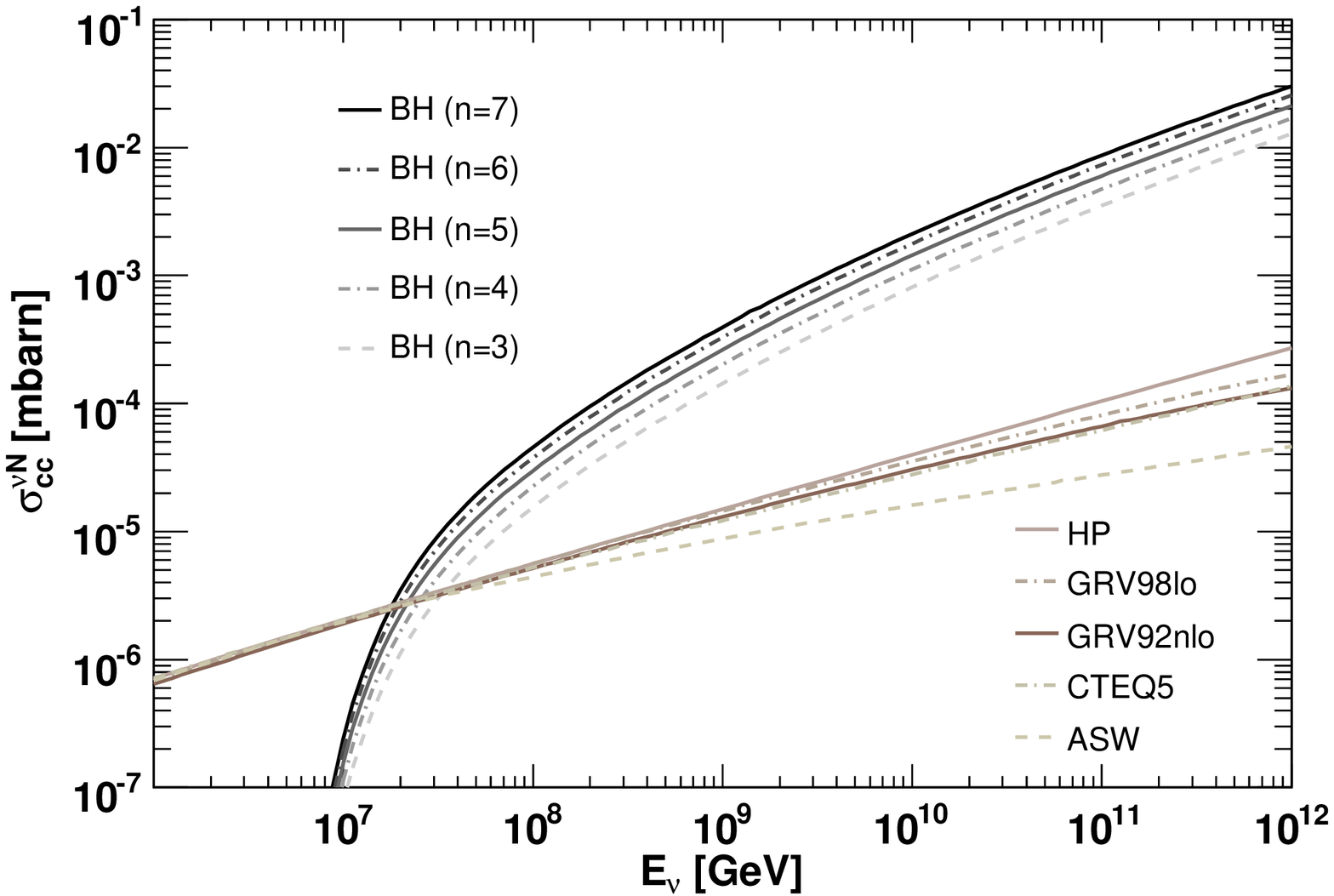}
\includegraphics[angle=0, width=0.48 \textwidth, height=5.99cm]{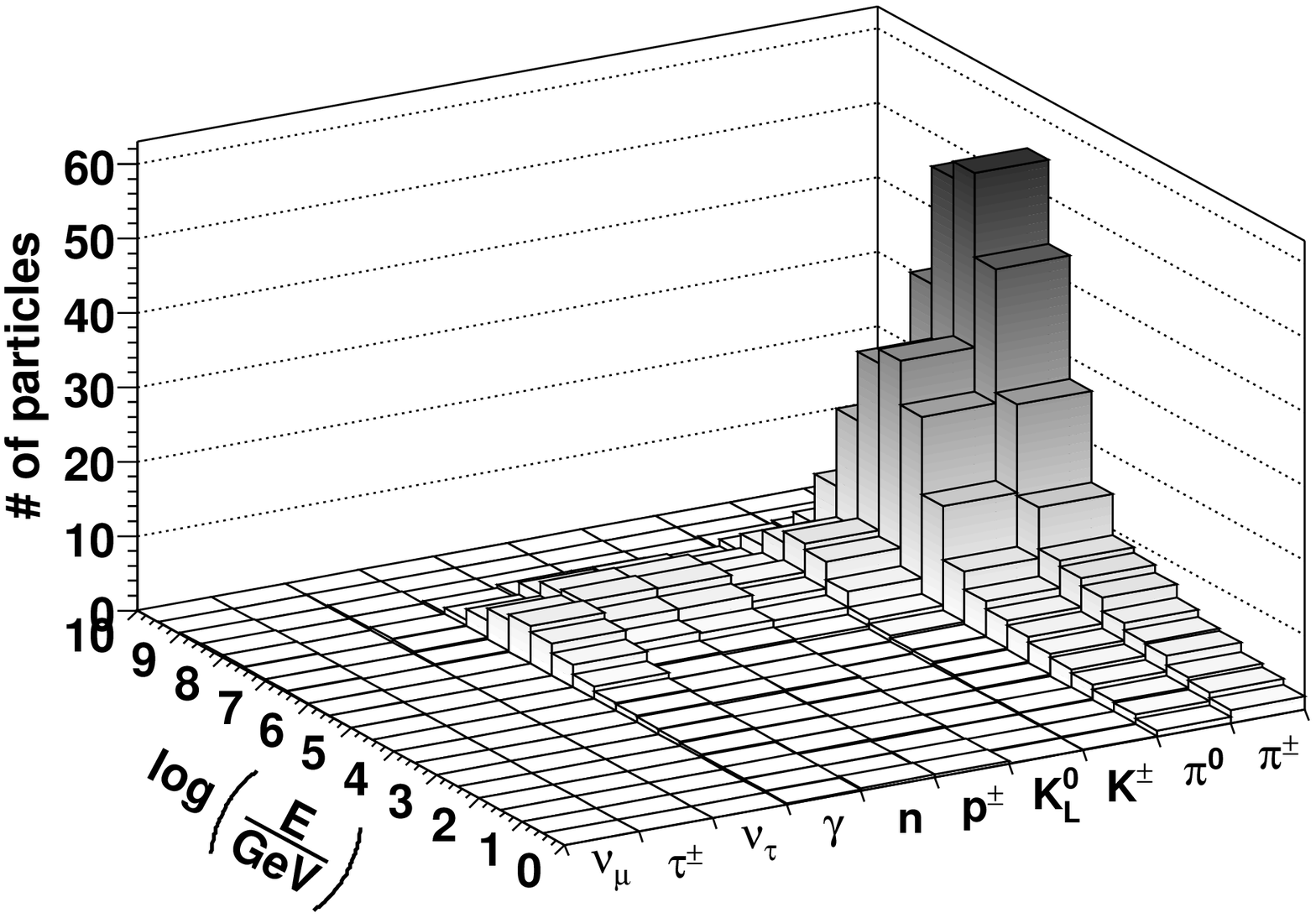}
\caption[Black hole cross section]{(Left panel) The BH formation cross section calculated for $x_{\min}=3$. For comparison the SM CC $\nu_{\tau} N$ cross section~\cite{remark} is shown; (right panel) particle spectrum of a BH decay with $E_{\nu_{\tau}} = 10^{19}\usk\electronvolt{}$, averaged over 100 events. The mean BH mass is $\langle M_\mathrm{BH} \rangle = 6730.8\usk\tera\electronvolt{}$, the multiplicity $\langle M \rangle = 614.4$. }
\label{fig:cs_bh}
\vspace{-15pt}
\end{figure*}

The decay and the resulting particle spectrum of BH was simulated using a modified version of the CHARYBDIS code~\cite{Harris:2003db}. The original version has been developed as an event generator for the production of microscopic BHs at the LHC.  One characteristic feature to be pointed out is the inclusion of the recently calculated \emph{grey-body} factors for BH production in extra dimensions~\cite{Kanti:2002ge}. Grey body factors account for the fact that particles have to be transmitted through a curved space-time outside the horizon, and result in a modified emission spectrum from that one of a perfect thermal black body, even in 4 dimensions. For sake of simplicity and due to the fact that the decay balding and Planck phase are not well understood, only the Hawking evaporation phase for a non-rotating BH is implemented. In CHARYBDIS the fragmentation and hadronization process of the radiated particles is realized by an interface to the generator PYTHIA~\cite{Sjostrand:2006za}.

There is no BH generator for collisions of neutrinos and nucleons available so far. Among other generators also CHARYBDIS was originally designed for $pp$ and $\bar{p}p$ collisions and motivated by collider experiments like the upcoming LHC. Modifications had to be introduced allowing for the study of neutrino-nucleon collisions. The neutrino has to be treated as a beam particle without substructure. This affects the calculation of the BH mass as well as the cross section, the handling of PDF and the initialization routine of PYHTIA. A Lorentz boost from the LAB system to the CM system before the collision and back after the decay was added to ensure the numerical stability even at highest energies. The implemented changes were verified by cross-checking the cross sections and distribution of the generated BH mass $M_\mathrm{BH}=\sqrt{xs}$ (which depends on the involved PDF) calculated in our cross section generator, CSGEN,
according to the cited literature against the results obtained with the modified version of CHARYBDIS.
The cross-section generator, CSGEN, is a self-written tool to calculate cross sections.
In CSGEN typical SM structure functions~\cite{remark} were
implemented as the parametrisation and  other  PDF included as the
LHAPDF~\cite{pdf} FORTRAN libraries.

All calculations were performed with the minimal black hole mass given by
$x_{\min} \equiv M_\mathrm{BH}^{\min} / M_\mathrm{D} = 3$ which corresponds to a lower cutoff where the semi-classical description of microscopic BHs is still valid, and the number of extra spacial dimensions set to $3 \leq n \leq 7$.

We determine the flux of neutrinos reaching the detector volume and initiating an extensive air shower (EAS). CSGEN is used to calculate cross sections, the distributions of involved kinematic variables and the tau energy loss for various interaction models and predictions~\cite{marco}. The data are input to a modified version of ANIS~\cite{Gazizov:2004va,Gora:2007nh} to simulate the propagation of incident neutrinos towards the detector and calculate the vertices of initiated air showers.

The probability to detect neutrinos and microscopic BH by the SD array of the Auger Observatory is done by means of the packages PYTHIA and CHARYBDIS, which are used to simulate the production of secondary particles in neutrino-nucleon interactions that eventually initiate an EAS. These particles are input to the air shower simulation software AIRES~\cite{aires} to create shower profiles and footprints, which are then analyzed with the Auger \Offline software framework~\cite{Argiro:2007qg} in order to determine the detector response and identification efficiency.

\section{Results}
The actual BH formation cross section for neutrino nucleon collisions calculated with CSGEN is shown in Fig.~\ref{fig:cs_bh} (left panel) for different assumptions on the number of additional spacial dimensions $n$. When the CM energy reaches values high enough to form a BH with the minimum mass $M_\mathrm{BH}^{\min}$ (which occurs between $E_{\nu}=10^{15}\usk\electronvolt{}$ and $E_{\nu}=10^{16}\usk\electronvolt{}$), the BH cross section rises rapidly, exceeding the SM cross sections by about two orders of magnitude at the highest energies. As an example the averaged spectrum of a BH decay at $E_{\nu_{\tau}} = 10^{19}\usk\electronvolt{}$ is shown in Fig.~\ref{fig:cs_bh} (right panel). It is evident that the secondaries consist mainly of charged and neutral pions, and kaons.

For the scenario of microscopic BH production, the neutrino interaction length in air is still larger than the atmospheric depth, so that BH showers can be initiated deep in the atmosphere and hence can be distinguished from hadronic cosmic rays in the same way like neutrino induced showers, i.e.\ looking for inclined young showers~\cite{Abraham:2008zz,jaime}.

Neutrinos are able to penetrate deep into the atmosphere before interacting and  generating a \emph{young shower} close to the detector as opposed to \emph{old showers} of hadronic or photon origin shortly after entering the atmosphere. At large zenith angles the purely electromagnetic part of such old showers is usually absorbed within the first $2000\usk\gram\usk\centi\metre^2$. Practically only the high energy muons reach the ground, especially in inclined showers. This results in a thin and flat shower front which generates a short detector signal, lasting only a few ten nanoseconds. Young showers however reach the ground with a significant electromagnetic component still existent, showing a curved and thick shower front at ground that leads to broad signals with durations of up to a few microseconds. Together with the time information of the particles detected at ground and the elongated shape of the footprint, young inclined showers can be identified and their origin eventually attributed to neutrinos. The larger the considered zenith angle, the more pronounced are these features.

\begin{figure}[tbp!]
\centering
\includegraphics[angle=0, width=0.49\textwidth, height=6cm]{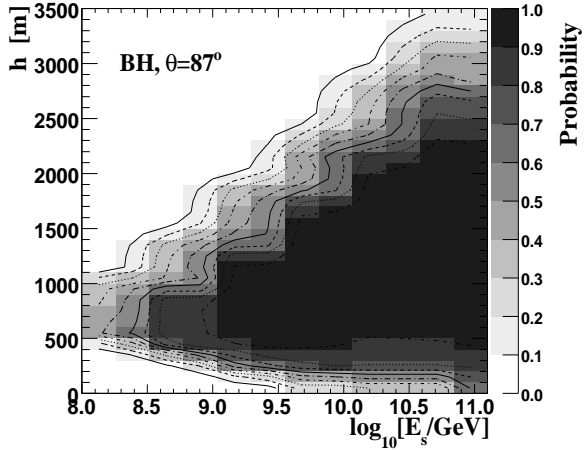}
\caption{The SD identification efficiency maps calculated for down-going $\nu_\mathrm{e}$   neutrinos producing microscopic BH. The efficiency is ploted as the function of   shower energy $E_\mathrm{s}$ and the injection height $h$. }
\label{eff}
\vspace{-15pt}
\end{figure}

To calculate the expected event rate from microscopic BHs we have defined a set of cuts where the number of events passing is maximal while the background contamination (due to hadron-induced showers) is kept minimal. A  method similar  to the one presented in~\cite{Abraham:2008zz} was used.  Applying the cuts to simulated neutrino and microscopic BH showers yield to the neutrino/BH identification efficiency, which is defined as the ratio of showers triggering the detector and passing the cuts over the total number of simulated AIRES showers. In Fig.~\ref{eff} an example of identification efficiencies is shown.

Once the identification is known the event rate can be calculated.  The total observable rates (number of expected events) were calculated according to $N=\Delta T \times \int_{E_\mathrm{th}}^{E_{\max}} A(E_{\nu})\times\Phi(E_{\nu})\times dE$, where $\Phi(E_{\nu})$ is the isotropic $\nu$-flux, $\Delta T$ the observation time and $A(E_{\nu})$ the acceptance for a given initial neutrino energy, $E_{\nu}$.
\begin{table}[b!]
\vspace{-10pt}
\caption{The expected event rates and the ratio between up- and down-going event     rates for different neutrino flux and cross sections.}
\centering
\begin{tabular}{ l l c c c }
\hline
\hline
 Scenario& flux $\Phi$ & rate 1/yr & rate 1/yr \\
  &  & up-g. & down-g. &  ratio $R$ \\
\hline
GRV92nlo & WB  & 0.44 & 0.22  & 2.0 \\
                                & GZK & 0.12 & 0.05  & 2.5 \\
                                & TD  & 0.49 & 0.32  & 1.6 \\
\hline
           HP & WB  & 0.23 & 0.26 & 0.9 \\
                    & GZK & 0.07 & 0.05 & 1.2 \\
                    & TD  & 0.25 & 0.39 & 0.7 \\
\hline
                 ASW & WB  & 0.54 & 0.13 & 4.2 \\
                     & GZK & 0.15 & 0.03 & 4.8 \\
                     & TD  & 0.61 & 0.17 & 3.6 \\

\hline
BH, n=5 & WB  & 0.02 & 11.30 & $\approx \frac{1}{500}$  \\
                         & GZK & 0.01 & 2.01  & $\approx \frac{1}{300}$  \\
                         & TD  & 0.02 & 19.31 & $\approx \frac{1}{1000}$ \\
\hline
\hline
\end{tabular}
\label{tab:ratio}
\end{table}

In Tab.~\ref{tab:ratio} the rates (number of events per year), for different injected $\nu$-fluxes are listed.  The rates labeled with ``WB'' are obtained for the Waxman-Bahcall bound~\cite{Waxman:1998yy}.  Other rates are calculated for the GZK flux~\cite{Engel:2001hd} and a flux due to Topological Defects (TD)~\cite{Bhattacharjee:1998qc}.  In addition in Tab.~\ref{tab:ratio} the ratio $R$ between expected up- and down-going neutrino events is presented for different interaction models. One finds that this ratio is indeed sensitive to the underlying neutrino-nucleon cross section. A higher cross section results in a decrease of up-going and increase of down-going event rates, and vice versa. But given the fact that for a non-exotic interaction model more than ten years of data-taking (depending on the incoming neutrino flux) might be necessary to detect a single down-going neutrino, the statistical relevance of the measured data will at best allow to put a limit on the cross section or energy loss model.  In the case of BH production the picture looks different: the non-observation of up-going neutrinos in conjunction with a high rate of down-going inclined air showers, initiated deep in the atmosphere, would be a strong indicator of physics beyond the SM.



\section{Conclusions}
A complete MC simulation chain to study the microscopic BH-induced showers has been presented.  Our calculations show a significant deviation of the expected event rate in comparison to the rate calculated by SM predictions. The non observation of up-going neutrinos by the Pierre Auger Observatory in conjunction with a high rate of down-going neutrino-induced showers, would be a strong indication of physics beyond the SM.
\section{Acknowledgment}
The authors gratefully acknowledge the fruitful discussions with our Pierre Auger collaborators, in particular we thank  J.~Alvarez-Mu\~niz for his strong interest. We are very much in dept to the authors of the Auger \Offline software used for the analysis.
We gratefully acknowledge the financial support by the HHNG-128 grant of the Helmholtz association and the Ministry of Science and Higher Education under Grant 2008 No. NN202 127235.


\end{document}